\documentclass[page-classic]{epl2} 
\usepackage{amsmath}

\title{Quantum oscillations in the linear chain of coupled orbits: the organic metal with two cation layers
$\theta$-(ET)$_4$CoBr$_4$(C$_6$H$_4$Cl$_2$)}
\shorttitle{Quantum oscillations in the linear chain of coupled orbits} 

\author{A. Audouard\inst{1} \and J.-Y. Fortin\inst{2} \and D. Vignolles\inst{1} \and R. B. Lyubovskii\inst{3} \and L. Drigo \inst{1} \and F.~Duc\inst{1} \and G. V. Shilov\inst{3} G. Ballon\inst{1} \and E. I. Zhilyaeva\inst{3} \and R. N. Lyubovskaya\inst{3} \and E. Canadell\inst{4}}
\shortauthor{A. Audouard \etal}

\institute{
  \inst{1} Laboratoire National des Champs Magn\'{e}tiques
Intenses (UPR 3228 CNRS, INSA, UJF, UPS) 143 avenue de Rangueil,
F-31400 Toulouse, France.\\
  \inst{2} Institut Jean Lamour, D\'epartement de Physique de la Mati\`ere et des Mat\'eriaux,
Groupe de Physique Statistique, CNRS-Nancy-Universit\'e
BP 70239 F-54506 Vandoeuvre les Nancy C\'edex, France.\\
  \inst{3} Institute of Problems of Chemical Physics, RAS, 142432 Chernogolovka, MD, Russia.\\
  \inst{4} Institut de Ci\`{e}ncia de Materials de Barcelona, CSIC, Campus de la UAB, 08193, Bellaterra, Spain.
}


\abstract{
Analytical formulae for de Haas-van Alphen (dHvA) oscillations in linear chain of coupled two-dimensional (2D) orbits
(Pippard's model) are derived systematically taking into account the chemical potential oscillations in magnetic
field. Although corrective terms are observed, basic ($\alpha$) and magnetic breakdown-induced
($\beta$ and 2$\beta$ - $\alpha$) orbits can be accounted for by the Lifshits-Kosevich (LK) and Falicov-Stachowiak
semiclassical models in the explored field and temperature ranges. In contrast, the 'forbidden orbit'
$\beta$ - $\alpha$ amplitude is described by a non-LK equation involving a product of two classical orbit
amplitudes. Furthermore, strongly non-monotonic field and
temperature dependence may be observed for the second harmonics of basic frequencies such as 2$\alpha$ and
the magnetic breakdown orbit $\beta$ + $\alpha$, depending on the value of the spin damping factors.
These features are in agreement with the dHvA oscillation spectra of the strongly 2D organic
metal $\theta$-(ET)$_4$CoBr$_4$(C$_6$H$_4$Cl$_2$).}

\begin{document}

\maketitle

\section{\label{sec:Intro}Introduction}

Quasi-two-dimensional (q-2D) multiband organic compounds (ET)$_2$X  (where ET stands for the
bis-ethylenedithio-tetrathiafulvalene molecule and X is a monovalent anion) exhibit very rich phase diagrams with ground states ranging from antiferromagnetic insulator to Fermi liquid. In particular, superconducting and magnetic ordered phases are observed close to each other which is reminiscent of cuprates and iron-based superconductors (for a recent review, see \cite{Ha11} and references therein). These compounds share the same Fermi surface (FS) topology which is an illustration of the linear chain of coupled orbits introduced
by Pippard in the early sixties \cite{Pi62}. Such FS is liable to give rise to a network of
orbits coupled by magnetic breakdown (MB) in large magnetic fields. Quantum oscillations in such systems have
been extensively studied during the two last decades (for a review, see $e.g.$ \cite{Ka04} and references therein).
The main feature of the oscillation spectra of these compounds is
the presence of frequency combinations that are forbidden in the framework of the semiclassical model of
Falicov-Stachowiak \cite{Sh84}. Theoretical studies have demonstrated that these Fourier components result
from both the formation of Landau bands due to coherent MB, instead of discrete levels, and the oscillation
of the chemical potential enabled by the 2D character of the FS. However, easy to handle analytic tools necessary
to quantitatively account for the data are still lacking. As a matter of fact, comprehensive theoretical calculations taking into account coherent MB and oscillation of the chemical potential have been reported in Ref.~\cite{Gv04} and
implemented in de Haas-van Alphen (dHvA) data of the above mentioned compound. Nevertheless, as pointed out in
Ref.~\cite{Gv04}, equations are complex and data could be analyzed only numerically.

The aim of this paper is therefore to provide analytical formulae accounting for quantum oscillation spectra of
these systems. In the first step, analytical calculation of the Fourier amplitude of the various dHvA oscillations
predicted for the FS illustrating the Pippard's model is reported. In the second step, the FS of the recently
synthesized strongly 2D charge transfer salt $\theta$-(ET)$_4$CoBr$_4$(C$_6$H$_4$Cl$_2$) \cite{Sh11} is presented
and, in the third step, analysis of the observed field- and temperature-dependent quantum oscillation spectra are
reported. An excellent agreement with the calculations is obtained.

\section{\label{sec:Experimental}Experimental}

\begin{figure}                                                    
\centering
\includegraphics[width=0.9\columnwidth,clip,angle=0]{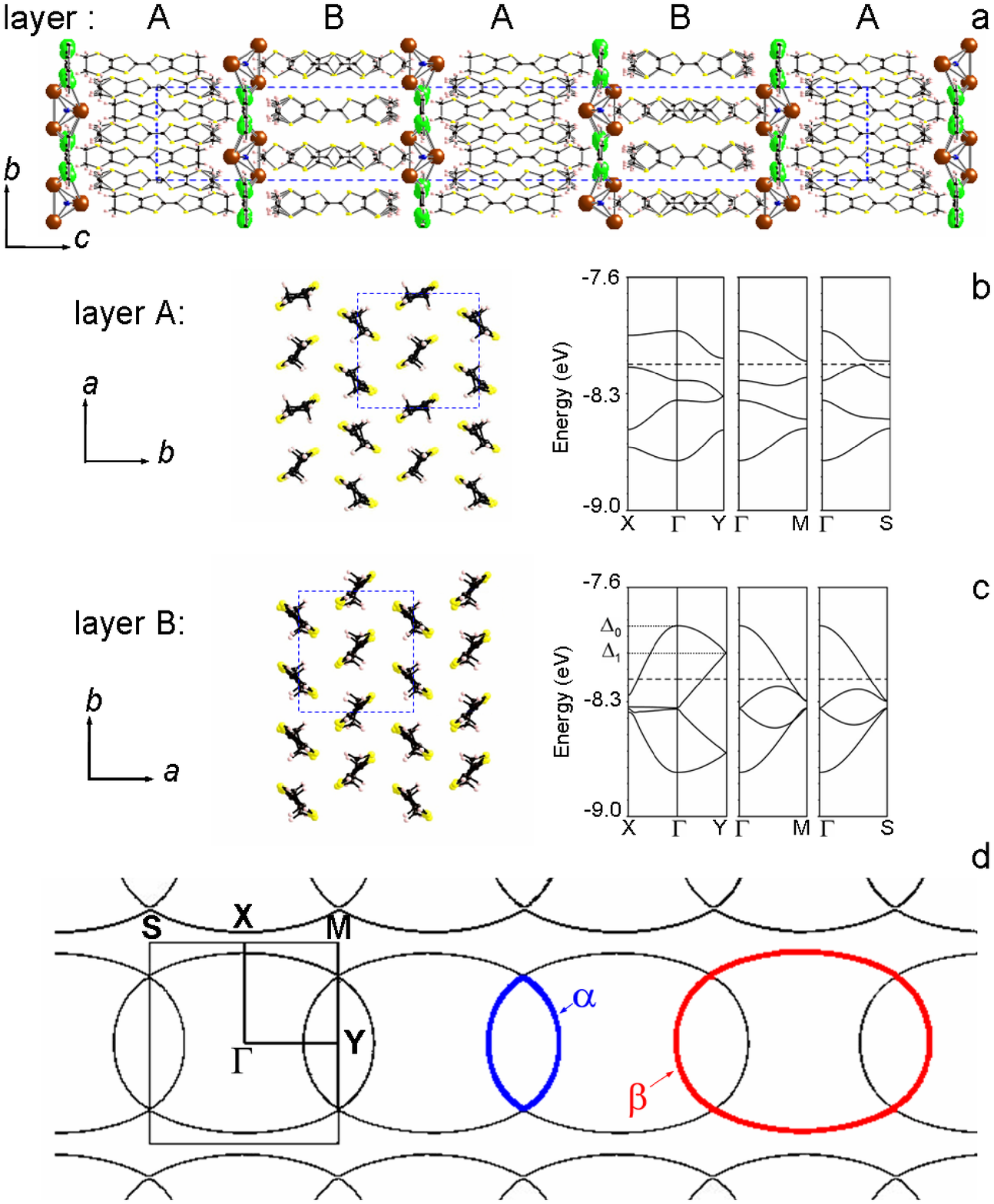} 
\caption{\label{Fig:Structure}(Colour on-line) (a) Crystal structure of $\theta$-(ET)$_4$CoBr$_4$(C$_6$H$_4$Cl$_2$)
projected along the $a$ direction. Two different cation layers labeled A and B, respectively, are evidenced.
Atomic arrangement of the ET molecules plane and band structure calculations relevant to (b) insulating layer A and (c) metallic layer B. (d) Fermi surface  relevant to layer B. Dashed line marks the top of the highest occupied band and the Fermi level in (b) and (c), respectively. Labels $\Delta_0$ and $\Delta_1$ stand for the band extrema. $\Gamma$, X, Y, M and S refer to (0, 0), ($a^*$/2, 0), (0, $b^*$/2), ($a^*$/2, $b^*$/2) and ($a^*$/2, -$b^*$/2), respectively.}
\end{figure}

The studied crystals were synthesized by electrocrystallization technique as reported in Ref.~\cite{Sh11}. They are
elongated hexagonal platelets with approximate dimensions 1.5 $\times$ 0.5 $\times$
0.04~mm$^3$. X-ray diffraction data were collected both at room temperature in Chernogolovka with a KM-4 single-crystal diffractometer (Kuma Diffraction) and at 100 K with an Xcalibur diffractometer (Oxford Diffraction) at the Laboratoire de Chimie de Coordination in Toulouse. Tight-binding band structure calculations were based upon the
effective one-electron Hamiltonian of the extended H\"{u}ckel method \cite{Wh78}, as reported $e.g.$ in
Ref.~\cite{Au11}. Magnetic torque and magnetoresistance were measured in pulsed magnetic fields of up to 55 T with a pulse decay duration of 0.32 s. Magnetic torque measurements were performed with a commercial piezoresistive microcantilever \cite{Oh02} in the temperature range from 1.4 K to 4.2 K. The size of the sample studied was approximately 0.12 $\times$ 0.1 $\times$ 0.04~mm$^3$. Variations of the cantilever piezoresistance was measured with a Wheatstone bridge with an $ac$ excitation at a frequency of 63 kHz. The angle between the normal to the conducting plane and the magnetic field direction was $\theta$ = 7$^{\circ}$. Magnetoresistance was measured thanks to a contactless radio frequency measurement technique based on tunnel
diode oscillator as reported in Ref. \cite{Dr10}. A pair of counter-wound coils made with copper wire of
50 $\mu$m in diameter was used. As pointed out in \cite{Oh04}, in the case where the interlayer resistivity is
much larger than the in-plane  resistivity, the contribution of the former parameter to the measured signal is
negligible. The time evolution of the oscillator frequency during the field pulse is deduced from successive short-term
fast Fourier transforms of the raw signal. In contrast, analysis of the oscillatory magnetoresistance and magnetic
torque is based on discrete Fourier transforms of the field-dependent data, calculated with a Blackman window after
removal of a non-oscillating background. Zero-field interlayer and in-plane 4-point resistance was measured with 1 $\mu$A $ac$ current at 77 Hz. Electrical contacts were made using annealed platinum wires of 20 $\mu$m in diameter glued with graphite paste.

\section{\label{sec:calculation}Calculation of de Haas-van Alphen amplitudes}
We consider a two-band system with band extrema $\Delta_{0(1)}$ (see Fig. \ref{Fig:Structure}(c)) and effective
masses $m_{0(1)}$. In the low temperature limit, the grand potential $\Omega$ of a two-dimensional slab with
area $\mathcal{A}$ is given by a series of harmonics $p$ of classical frequencies $F_\eta$:

\begin{eqnarray}
\nonumber
\label{Eq:Omega}\frac{\Phi_0}{\mathcal{A}}\frac{u_0}{k_B}\Omega(T,\mu)=-\frac{m_0}{2}(\mu-\Delta_0)^2
-\frac{m_1}{2}(\mu-\Delta_1)^2
\\
+\frac{(B\cos\theta)^2}{2}\sum_{p\geq1}\sum_{\eta}\frac{C_{\eta}}{\pi^2p^2m_{\eta}}
R_{\eta,p}(T,B)\cos(2\pi p\frac{F_{\eta}}{B\cos\theta}+p\varphi_{\eta}),
\end{eqnarray}

where $\Phi_0$ = $h$/$e$ is the quantum flux and $u_0$ = 2$\pi^2m_ek_B/e\hbar$ = 14.694 T/K.
Effective masses are expressed in $m_e$ units and chemical potential $\mu$ in Tesla. The index $\eta$ stands for all the possible classical orbits $\alpha$, $\beta$, $\beta$ + $\alpha$,
2$\beta$ - $\alpha$, $etc.$, excluding their harmonics, with
frequencies $F_{\eta}$ = $m_{\eta}$($\mu$ - $\Delta_{\eta}$) and effective masses $m_{\eta}$.
Frequencies $F_{\eta}$ depend on the chemical potential $\mu$ since they represent a filling factor
corresponding to classical orbits.
Coefficient $C_{\eta}$ is the symmetry factor of the orbit $\eta$. For example,
$C_{\alpha}=C_{\beta}=C_{2\beta-\alpha}=1$ and $C_{\beta+\alpha}=2$. $\varphi_{\eta}$
is the phase of the orbit $\eta$, $i.e.$ $\pi/2$ times the number of extrema of the orbit in the $k_x$ or $k_y$
direction, for example, $\varphi_{\alpha}=\varphi_{\beta}=\varphi_{2\beta-\alpha}=\pi$ and
$\varphi_{\beta+\alpha}=2\pi$. These phases are important for determining the sign of amplitudes.
In particular, we define two fundamental orbits with frequencies, $F_{\alpha}$ = $m_1$($\mu$ - $\Delta_{1}$)
($i.e.$  $m_{\alpha}$ = $m_1$) and $F_{\beta}$ = $m_0$($\mu$ - $\Delta_{0}$) + $m_1$($\mu$ - $\Delta_{1}$) which
can be written as $F_{\beta}$ = $m_{\beta}$($\mu$ - $\Delta_{\beta}$) (see Fig.~\ref{Fig:Structure}(d)). Therefore, $m_0$ + $m_1$ is identified to
the effective mass $m_{\beta}$ of the MB orbit $\beta$. The damping factors can be expressed as
$R_{\eta,p}(T,B)$ = $R^{T}_{\eta,p}R^{D}_{\eta,p}R^{MB}_{\eta,p}R^{s}_{\eta,p}$ where the thermal, Dingle,
MB and spin damping factors are given by:

\begin{eqnarray}
\label{Eq:RT}R^{T}_{\eta,p} = \frac{u_0Tpm_{\eta}}{B\cos\theta\sinh(\frac{u_0Tpm_{\eta}}{B\cos\theta})} \\
\label{Eq:RD}R^{D}_{\eta,p} = \exp(-\frac{u_0T_Dpm_{\eta}}{B\cos\theta}) \\
\label{Eq:RMB}R^{MB}_{\eta,p} = (ip_0)^{n^t_{\eta}}(q_0)^{n^r_{\eta}}\\
\label{Eq:Rs}R^{s}_{\eta,p} = \cos(\pi p\frac{g^*m_{\eta}}{2\cos\theta}),
\end{eqnarray}

respectively. $T_D$ is the Dingle temperature ($T_{D}$ = $\hbar$/2$\pi$$k_B\tau$ where $\tau$ is the
relaxation time).
Integers $n^t_{\eta}$ and $n^r_{\eta}$ are the (even) number of MB tunnelings and reflections, respectively,
encountered by the quasiparticle along its closed trajectory. The MB tunneling and reflection probabilities are
given by $p_0$ = $\exp(-B_0/2B)$ and $q_0^2$ = 1 - $p_0^2$, respectively, where $B_0$ is the MB field.
$g^*$ is the effective Land\'{e} factor\footnote{FS warping can be accounted for by the additional damping factor J$_0$(4$\pi pm_{\eta}\tau_{\perp}$/$e\hbar B$), where $\tau_{\perp}$ is the interlayer transfer integral \cite{Ch01}. In most q-2D organic metals, this factor is close to 1. }.
The electron density per surface area $n_e$ is given by d$\Omega$/d$\mu$ = -$n_e$.
In zero-field, Eq.~\ref{Eq:Omega} yields $n_e$ = ($m_0$ + $m_1$)$\mu_0$ - $m_0\Delta_0$ - $m_1\Delta_1$.
In presence of a magnetic field, $\mu$ satisfies the following implicit equation:

\begin{eqnarray}
\label{Eq:mu}\mu=\mu_0-\frac{B\cos\theta}{m_{\beta}}\sum_{p\geq1}\sum_{\eta}\frac{1}{\pi p}
C_{\eta}R_{\eta,p}(T,B)\sin(2\pi p \frac{F_{\eta}}{B\cos\theta}+p\varphi_{\eta})
\end{eqnarray}

The oscillating part of the magnetization can be computed systematically by inserting Eq.~\ref{Eq:mu} in Eq.~\ref{Eq:Omega} and deriving the resulting expression with respect to the magnetic field. Direct numerical resolution of this implicit equation could be done. Besides, a more user-friendly controlled expansion in term of damping factors $R_{\eta,p}$ can be derived systematically at any possible order. Keeping successive terms in powers of $R_{\eta,p}$ up to the second order,
yields analytically the following expression for the oscillating part of the magnetization:

\begin{eqnarray}
\nonumber
\label{Eq:mosc}m_{osc}&=&-\sum_{\eta}\sum_{p\ge 1}\frac{F_{\eta}C_{\eta}}{\pi p m_{\eta}}
R_{\eta,p}(T,B)\sin\left ( 2\pi p \frac{F_{\eta}}{B\cos\theta}+p\varphi_{\eta} \right )
\\ \nonumber
&+&\sum_{\eta,\eta'}\sum_{p,p'\ge 1}\frac{F_{\eta}C_{\eta}C_{\eta'}}{\pi p' m_{\beta}}
R_{\eta,p}(T,B)R_{\eta',p'}(T,B)
\left [
\sin\left ( 2\pi \frac{pF_{\eta}+p'F_{\eta'}}{B\cos\theta}+p\varphi_{\eta}+p'\varphi_{\eta'} \right )
\right .
\\
&-&\left .\sin\left ( 2\pi \frac{pF_{\eta}-p'F_{\eta'}}{B\cos\theta}+p\varphi_{\eta}-p'\varphi_{\eta'} \right )
\right ]+\cdots
\end{eqnarray}

where frequencies $F_{\eta}$ are evaluated at $\mu$ = $\mu_0$, namely,
$F_{\eta}=m_{\eta}(\mu_0-\Delta_{\eta})$. Eq.~\ref{Eq:mosc} can be written as a sum of periodic functions
with frequencies $F_i$ and Fourier amplitudes $A_i$ where the index $i$ stands for either classical orbits
$\eta$ or 'forbidden' orbits such as $\beta$ - $\alpha$. Relevant examples for experimental analysis are given below:

\begin{eqnarray}
\label{Eq:Aalpha}A_{\alpha}&=&\frac{F_{\alpha}}{\pi m_{\alpha}}R_{\alpha,1}
+\frac{F_{\alpha}}{2\pi m_{\beta}}R_{\alpha,1}R_{\alpha,2}
+\frac{F_{\alpha}}{6\pi m_{\beta}}R_{\alpha,2}R_{\alpha,3}
+\frac{2F_{\alpha}}{\pi m_{\beta}}R_{\beta,1}R_{\alpha+\beta,1},
\\
\label{Eq:A2alpha}A_{2\alpha}&=&-\frac{F_{\alpha}}{2\pi m_{\alpha}}R_{\alpha,2}+
\frac{F_{\alpha}}{\pi m_{\beta}}R_{\alpha,1}^2-\frac{2F_{\alpha}}{3\pi m_{\beta}}R_{\alpha,1}R_{\alpha,3},
\\
\label{Eq:Abeta}A_{\beta}&=&\frac{F_{\beta}}{\pi m_{\beta}}R_{\beta,1}
+\frac{F_{\beta}}{2\pi m_{\beta}}R_{\beta,1}R_{\beta,2}
+\frac{F_{\beta}}{6\pi m_{\beta}}R_{\beta,2}R_{\beta,3}
+\frac{2F_{\beta}}{\pi m_{\beta}}R_{\beta,1}R_{\alpha+\beta,1},
\\
\label{Eq:A2beta}A_{2\beta}&=&-\frac{F_{\beta}}{2\pi m_{\beta}}R_{\beta,2}+
\frac{F_{\beta}}{\pi m_{\beta}}R_{\beta,1}^2-\frac{2F_{\beta}}{3\pi m_{\beta}}R_{\beta,1}R_{\beta,3},
\\
\label{Eq:Abeta-alpha}A_{\beta-\alpha}&=&-\frac{F_{\beta-\alpha}}{\pi m_{\beta}}R_{\alpha,1}R_{\beta,1}
-\frac{F_{\beta-\alpha}}{\pi m_{\beta}}R_{\alpha,2}R_{\alpha+\beta,1}
-\frac{F_{\beta-\alpha}}{\pi m_{\beta}}R_{\beta,2}R_{\alpha+\beta,1},
\\
\label{Eq:Abeta+alpha}A_{\beta+\alpha}&=&-\frac{2F_{\beta+\alpha}}{\pi m_{\beta+\alpha}}
R_{\beta+\alpha,1}+\frac{F_{\beta+\alpha}}{\pi m_{\beta}}R_{\alpha,1}R_{\beta,1},
\\
\label{Eq:A2beta-alpha}A_{2\beta-\alpha}&=&\frac{F_{2\beta-\alpha}}{\pi m_{2\beta-\alpha}}
R_{2\beta-\alpha,1}+
\frac{F_{2\beta-\alpha}}{2\pi m_{\beta}}
R_{\alpha,1}R_{\beta,2}+
\frac{F_{2\beta-\alpha}}{6\pi m_{\beta}}
R_{\alpha,3}R_{\alpha+\beta,2}.
\end{eqnarray}

\begin{figure}                                                    
\centering
\includegraphics[width=0.9\columnwidth,clip,angle=0]{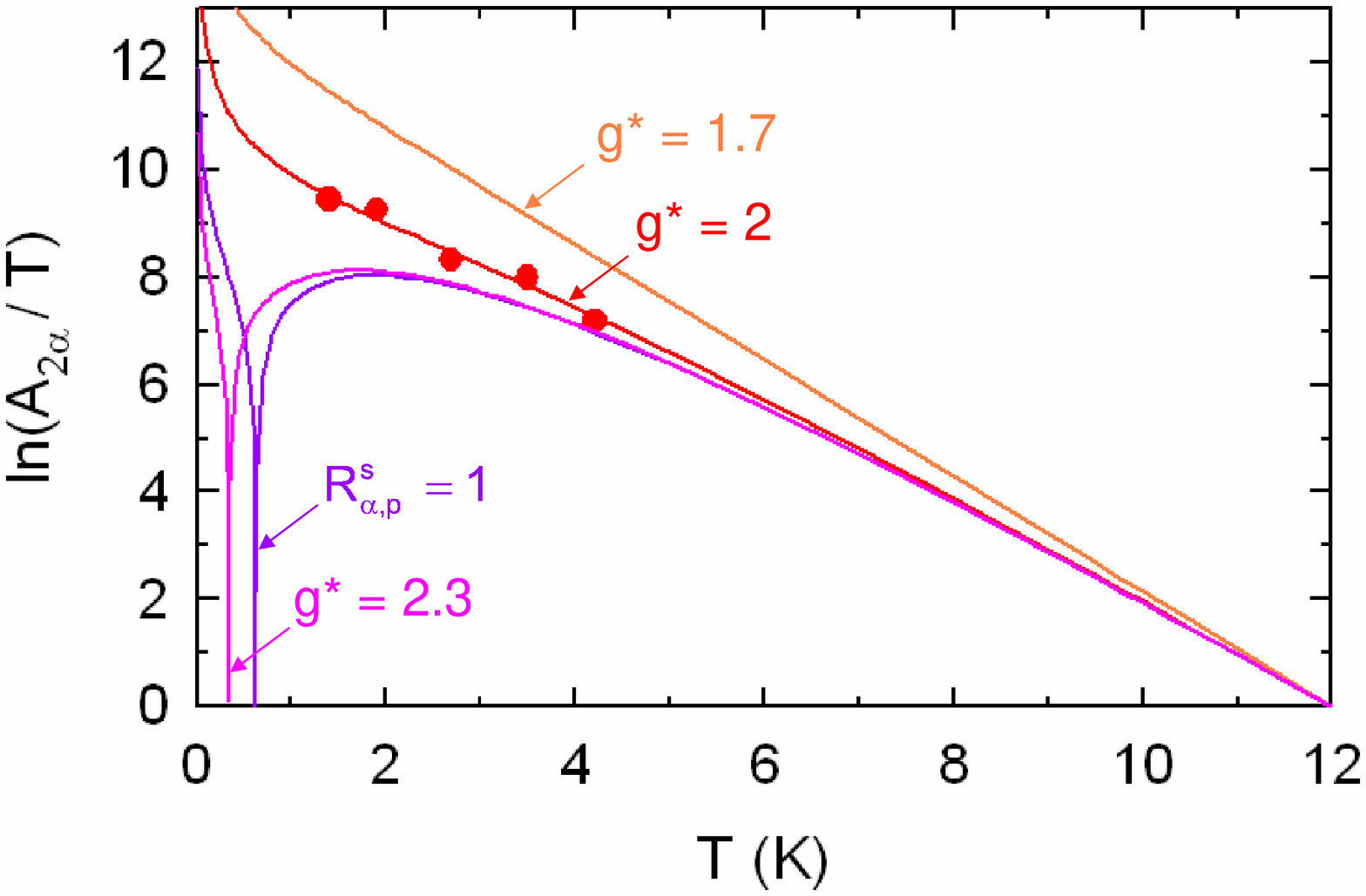}
\caption{\label{Fig:mass_plot_2a} (Colour on-line)  Temperature dependence (mass plot) of the amplitude of
the component 2$\alpha$. Symbols are data deduced from torque measurements of  Fig.~\ref{Fig:data_TF}(a)
at 48.3 T. Solid lines are deduced from Eq.~\ref{Eq:A2alpha} with $m_\alpha$ = 1.81, $m_\beta$ = 3.52,
$T_D$ = 0.79 K and $B_0$ = 35 T. Spin damping factor (see Eq.~\ref{Eq:Rs}) is either neglected ($R^s_p$ = 1, p = 1 to 3) or
accounted for by various $g^*$ values. All the data are normalized to their values at 12 K. }
\end{figure}

In the framework of the LK model, the amplitude of a Fourier component linked to a basic orbit involves only one
damping factor ($e.g.$ $A_{\alpha} \propto R_{\alpha}$). This is also the case for a MB orbit, the effective mass
of which is given by the Falicov-Stachowiak model ($e.g.$ $m_{2\beta-\alpha}$ = 2$m_{\beta}$ - $m_{\alpha}$)
\cite{Sh84}. According to this statement, the leading term of Eqs.~\ref{Eq:Aalpha} and \ref{Eq:Abeta} which accounts for the $\alpha$ and $\beta$ components, respectively, reproduces the LK equation. Furthermore, deviations from the LK behaviour due to the high order terms involving products of damping factors are only significant in the low temperature and high field ranges. This statement also stands for 2$\beta$-$\alpha$ since $R_{2\beta-\alpha}$ is significantly higher than the product $R_{\alpha,1}R_{\beta,2}$ (see Eq.~\ref{Eq:A2beta-alpha}). As a consequence, the
Falicov-Stachowiak model should apply for this component as well. In contrast, algebraic sum of damping terms
in Eqs.~\ref{Eq:A2alpha} and~\ref{Eq:Abeta+alpha}, where minus signs accounts for $\pi$ dephasings, may
cancel at field and temperature values dependent on the effective masses, Dingle temperature, MB field, $etc.$
as displayed in Fig.~\ref{Fig:mass_plot_2a} relevant to 2$\alpha$. This point has already been reported for the
second harmonic of the basic orbits both for compensated \cite{Fo09} and un-compensated \cite{Fo05} metals. In
that respect, since algebraic sums are involved, care must be taken of the spin damping factor sign\footnote{Since the LK equation involves a single damping term, the absolute value of the spin damping factor
is considered in the case where data are analyzed on this basis. However, $\pi$ dephasing is observed on either
side of spin-zero angles as reported in \cite{Si00}.}. Indeed, the value of the spin damping factor, which is changed
through the value of $g^*$ in Fig.~\ref{Fig:mass_plot_2a}, can have a drastic effect on the field and temperature
dependence of the amplitude in this case. Finally, the Fourier amplitude of the 'forbidden orbit'
$\beta$ - $\alpha$ is accounted for by Eq.~\ref{Eq:Abeta-alpha}. Although a non-LK behaviour is evidenced, a monotonic field and temperature dependence is expected owing to the negligibly small value of the high order terms. The consistency of Eqs.~\ref{Eq:Aalpha} to~\ref{Eq:A2beta-alpha} with experimental data is examined
in the next section.

\section{\label{sec:Results}Results and discussion}

\begin{figure}                                                    
\centering
\includegraphics[width=0.9\columnwidth,clip,angle=0]{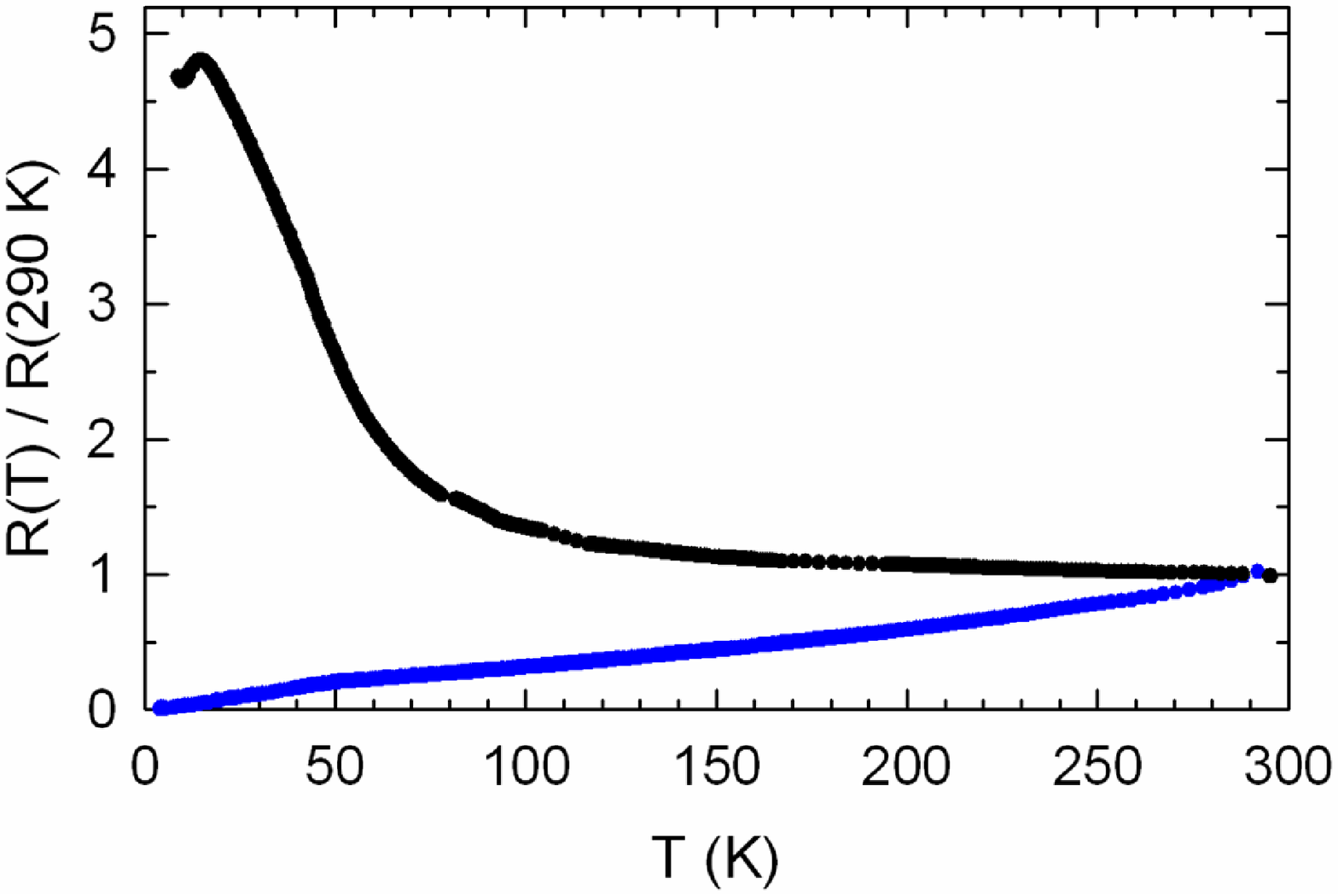}
\caption{\label{Fig:RT} (Colour on-line)  Temperature dependence of zero-field in-plane (blue lower curve) and
interlayer (black upper curve) resistance normalized to the values at 290 K. }
\end{figure}

Even though the organic metal $\theta$-(ET)$_4$CoBr$_4$(C$_6$H$_4$Cl$_2$) undergoes a phase transition at 343 K
\cite{Sh11}, no additional phase transition is revealed by X-ray diffraction below room temperature, at which crystals are synthesized, down to 100 K. As displayed in Fig.~\ref{Fig:Structure}, the central feature of the crystalline
structure at room temperature is that the unit cell has two different donors planes labeled A and B. According to band structure calculations, layer A is insulating while layer B is metallic. FS of layer B is composed of one closed tube centered at Y with an area of 18 $\%$ of the first Brillouin zone (FBZ) area (yielding $\alpha$ orbit in magnetic field) and two q-1D sheets with rather small gap in between. Therefore, it is expected that the MB orbit $\beta$, which corresponds to the hole tube centered at $\Gamma$ with an area equal to that of the FBZ is
observed in moderate fields, giving rise to the textbook linear chain of coupled orbits. In addition, since
layers A are insulating, conducting planes are strongly separated from each other which suggests a strongly
2D behaviour. In line with this statement, the room temperature resistivity anisotropy $\rho$(interlayer)/$\rho$
(in plane) is as large as 10$^4$~\cite{Sh11}. Furthermore, according to the data in Fig.~\ref{Fig:RT},
this anisotropy further increases up to $\rho$(interlayer)/$\rho$(in plane) =
10$^6$ as the temperature decreases down to liquid helium temperatures.

\begin{figure}                                                    
\centering
\includegraphics[width=0.9\columnwidth,clip,angle=0]{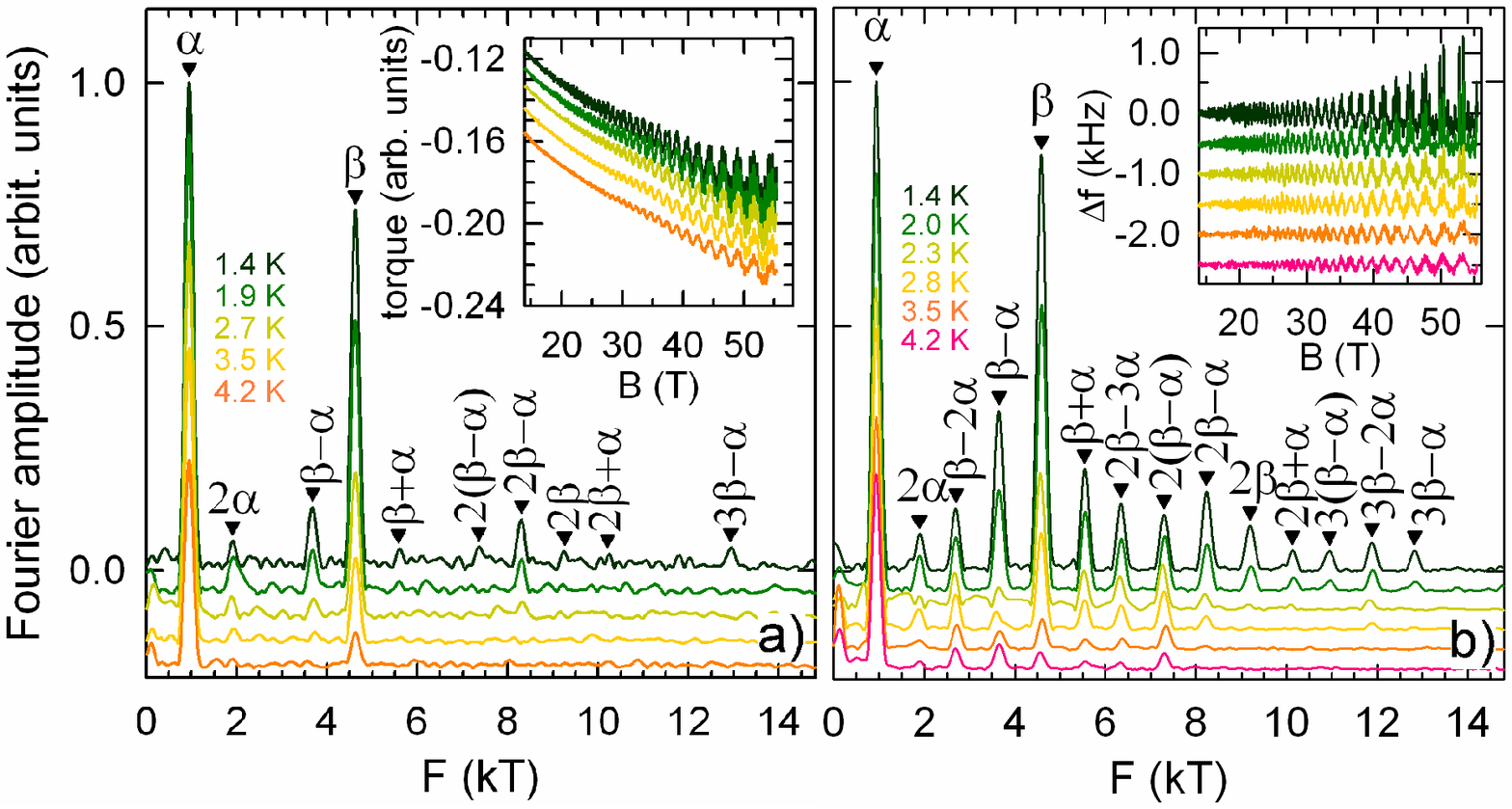}
\caption{\label{Fig:data_TF} (Colour on-line) Fourier analysis in the field range 35-55.3 T of (a) the magnetic
torque data (dHvA oscillations) and (b) the tunnel diode oscillator data (SdH oscillations) reported in the respective insets. Solid triangles are marks calculated with F$_{\alpha}(\theta = 0)$ = 0.944 kT
and F$_{\beta}(\theta = 0)$ = 4.60 kT. Data have been shifted down from each other by a constant amount.}
\end{figure}

\begin{figure}                                                    
\centering
\includegraphics[width=0.9\columnwidth,clip,angle=0]{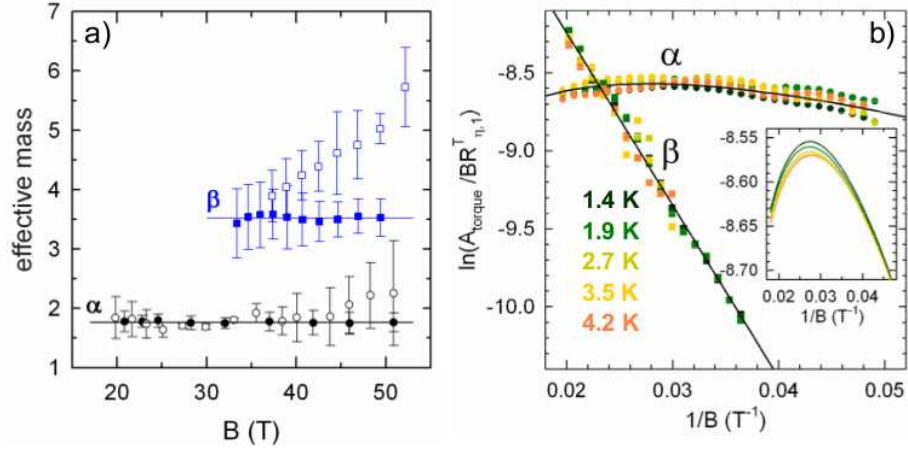}
\caption{\label{Fig:alpha_beta} (Colour on-line) (a) Effective masses vs. magnetic field for the $\alpha$ (circles) and $\beta$ (squares) oscillations deduced from best fits of Eq.~\ref{Eq:Aalpha} to the torque (solid symbols) and tunnel diode oscillator (open symbols) data. Horizontal lines stand for the mean values of the effective masses deduced from torque data ($m_{\alpha}$ = 1.81, $m_{\beta}$ = 3.52). (b) Dingle plot of the $\alpha$ and $\beta$
torque oscillations. $R^T_{\eta,1}$, where $\eta$ = $\alpha$, $\beta$, is calculated with $m_{\alpha}$ and
$m_{\beta}$ deduced from (a). Solid lines are best fits of Eq.~\ref{Eq:Aalpha} and~\ref{Eq:Abeta} to the data relevant to $\alpha$ and $\beta$, respectively, obtained with B$_0$ = 35 T and $T_D$ = 0.79 K  (see Eqs.~\ref{Eq:RT} to~\ref{Eq:RMB}). Inset displays an enlarged view of the fits for $\alpha$.}
\end{figure}

\begin{figure}                                                    
\centering
\includegraphics[width=0.9\columnwidth,clip,angle=0]{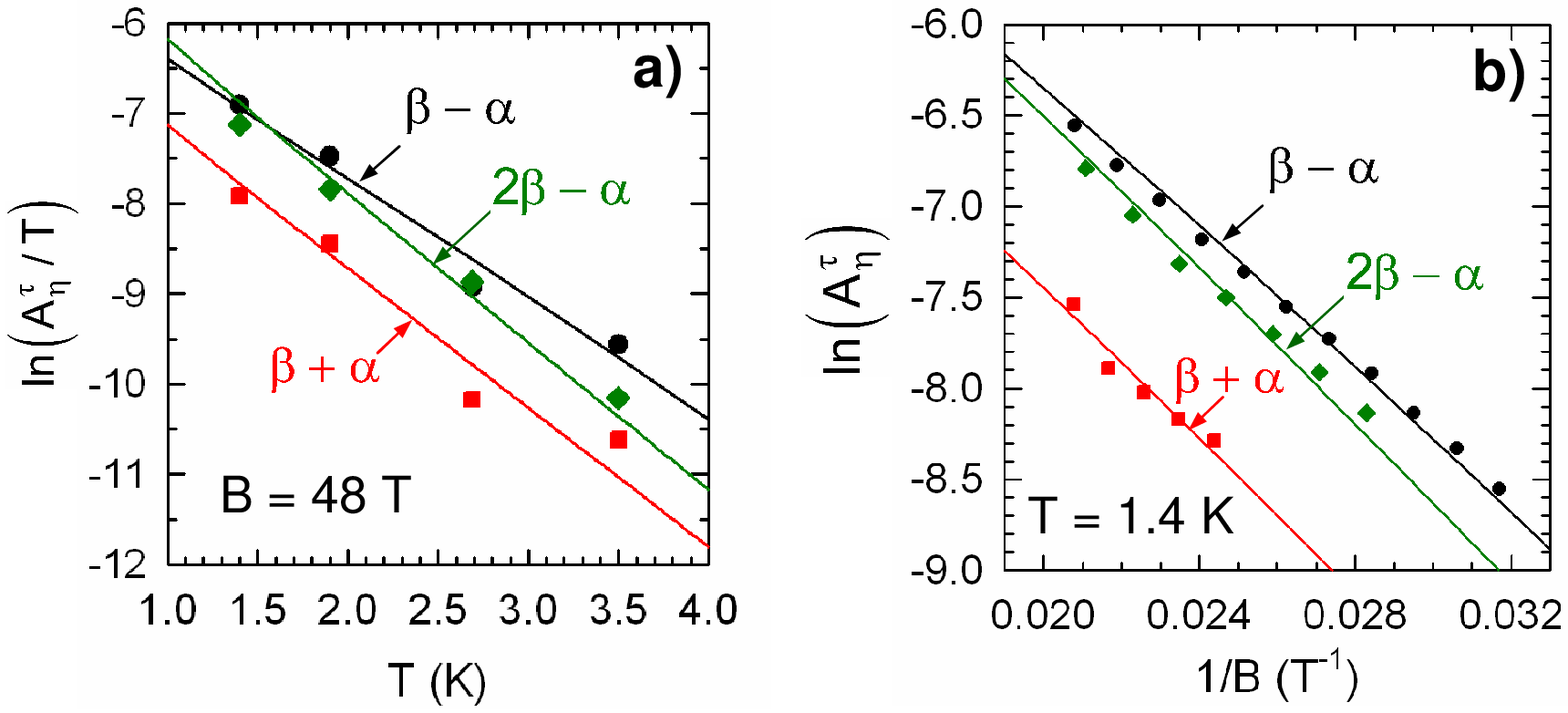}
\caption{\label{Fig:combinaisons} (Colour on-line) (a) Mass plots and (b) Dingle plots of $\beta$ - $\alpha$
(circles),  $\beta$ + $\alpha$ (squares) and 2$\beta$ - $\alpha$ (diamonds). Solid lines are bests fits of
Eqs.~\ref{Eq:Abeta-alpha},~\ref{Eq:Abeta+alpha} and~\ref{Eq:A2beta-alpha}, calculated with $m_{\alpha}$ = 1.81,
$m_{\beta}$ = 3.52, $B_0$ = 35 T, $T_D$ = 0.79 K and $g^*$ = 2.}
\end{figure}

Magnetic torque (dHvA oscillations) and tunnel diode oscillator [Shubnikov-de Haas (SdH) oscillations] data are displayed in  Fig.~\ref{Fig:data_TF}.
Besides Fourier components linked to the closed orbit $\alpha$ and to the MB orbit $\beta$, few frequency
combinations typical of the FS displayed in Fig.~\ref{Fig:Structure} are observed in the high field range of
the torque spectra. In contrast, numerous combinations are observed in SdH spectra. As an example, the Fourier
component 2$\beta$ - 3$\alpha$  is clearly detected in Fig.~\ref{Fig:data_TF}(b). This frequency, which to our knowledge is observed for the first time, could correspond to a quantum interferometer path. However, its amplitude is unexpectedly large. The frequencies linked to the $\beta$ and $\alpha$ orbits are $F_{\beta}$($\theta$ = 0) = 4600 $\pm$ 10 T, in agreement with the FBZ area at 100 K, and $F_{\alpha}$($\theta$ = 0) = 944 $\pm$ 4 T,
respectively. The latter value yields a cross section area of 20.6 \% of the FBZ area, in satisfactory agreement
with the band structure calculations of Fig.~\ref{Fig:Structure}. Field- and temperature-dependent Fourier
amplitudes can be studied at the light of the results given in the preceding section, keeping in mind that
dHvA amplitudes A$_{\eta}$ are related to torque oscillation amplitudes $A^{\tau}_{\eta}$ as $A_{\eta}
\propto A^{\tau}_{\eta}$/$B$tan($\theta$). Effective masses, derived from Eq.~\ref{Eq:Aalpha} at various
magnetic field values are field-independent: $m(\alpha)$ = 1.81 $\pm$ 0.05,  $m(\beta)$ = 3.52 $\pm$ 0.19
(see Fig.~\ref{Fig:alpha_beta}(a)). In agreement with results of Ref. \cite{Me00}, $m(\beta)$ is twice $m(\alpha)$
within the error bars. At odds with torque results, effective masses deduced from SdH data, in particular those
relevant to the $\beta$ component, definitely increase as the magnetic field increases. This behaviour indicates
that the above calculations, relevant to dHvA oscillations, cannot account for the SdH data at high field.

Focusing on dHvA oscillations, Dingle plots of $\alpha$ and $\beta$ can be obtained plotting
$A^{\tau}_{\eta}$/$BR^T_{\eta,1}$ where $R^T_{\eta,1}$ is given by Eq.~\ref{Eq:RT}, assuming the LK formalism accounts
for the data, i.e. the high orders terms of  Eq.~\ref{Eq:Aalpha} can be neglected. As reported in
Fig.~\ref{Fig:alpha_beta}, all the data relevant to the $\alpha$ and $\beta$ components lies, respectively,
on the same curve within the experimental scattering which suggests that the above assumption holds.
Dingle temperature and MB field can be derived for components which do not involve tunneling ($n^t_{\eta}$ = 0
in Eq.~\ref{Eq:Rs}). Although this is the case of $\alpha$, a very broad maximum is observed in
Fig.~\ref{Fig:alpha_beta}, leading to a very large uncertainty on the MB field. More specifically,
$B_0$ values ranging from 0 to 70 T can account for the data, in which case T$_D$ ranges from 1.4 to 0.5 K,
respectively. Assuming the Dingle temperature linked to the $\alpha$ orbit is the same as that of $\beta$
considerably reduces the uncertainty since the well defined slope $s_{\beta}$ = -($u_0T_Dm_{\beta}$+2$B_0$) of the Dingle plot for $\beta$ yield an additional relationship between T$_D$ and B$_0$. This assumption yields $B_0$ = 35 $\pm$ 5 T and $T_D$ = 0.79
$\pm$  0.10 K. With these values of effective masses, Dingle temperature and MB field, it is possible to check
whether or not the high order terms of Eq.~\ref{Eq:Aalpha} can be neglected. Inset
of Fig.~\ref{Fig:alpha_beta}(b) displays the field dependence of $A^{\tau}_{\alpha}$/$BR^T_{\alpha,1}$ calculated
for the highest field range covered by the experiments, assuming further that $g^*$ = 2\footnote{Effective Land\'{e} factor as low as $g^*$ = 1.5 $\div$ 1.6 are reported for the strongly
correlated compound $\kappa$-(ET)$_2$Cu(NCS)$_2$ \cite{Gv04}, nevertheless $g^*$ = 2 is consistent with
data of other ET salts \cite{Au05,Vi09}.}. Although the contribution of these factors rises as the field
increases, it remains smaller than the experimental scattering for both $\alpha$ and $\beta$.

The field and temperature dependence of the Fourier amplitude relevant to frequency combinations is displayed
in Fig.~\ref{Fig:combinaisons}. Since their amplitude is much smaller than that of the orbits $\alpha$ and $\beta$,
only mass plots at high field and Dingle analysis at the lowest explored temperature are considered. As already
discussed for the second harmonics 2$\alpha$ (see Fig.~\ref{Fig:mass_plot_2a}), amplitudes are strongly dependent
on the spin damping factors value. Solid lines in Fig.~\ref{Fig:combinaisons} are the best fits of
Eqs.~\ref{Eq:Abeta-alpha} to \ref{Eq:A2beta-alpha}, obtained with the same values of $m_{\alpha}$, $m_{\beta}$,
$B_0$, $T_D$ and  $g^*$ as above, with only a prefactor as a free parameter for each of the data: a very good
agreement is obtained, in particular for the 'forbidden orbit' $\beta$ - $\alpha$. It can be remarked that
the behaviour of both 2$\beta$ - $\alpha$ and $\beta$ + $\alpha$ are close to the predictions of the
Falicov-Stachowiak model. Whereas this feature is mainly due to the small value of $R_{\alpha,1}R_{\beta,2}$
compared to $R_{2\beta-\alpha,1}$ in the former case, it can be attributed to the peculiar set of values of the
spin damping factors ($R^s_{\alpha,1}R^s_{\beta,2}$/$R^s_{\beta + \alpha,1}$ = -0.22) in the latter case.

\section{\label{sec:Conclusion}Summary and conclusion}

Analytical formulae for dHvA oscillations relevant to a class of FS illustrating the Pippard's model and observed in many q-2D organic conductors have been derived systematically taking into account the chemical potential oscillations in magnetic field (Eqs.~\ref{Eq:Aalpha}
to ~\ref{Eq:A2beta-alpha}). Although high orders terms are observed, basic ($\alpha$) and MB-induced ($\beta$
and 2$\beta$ - $\alpha$) orbits can be accounted for by the LK and Falicov-Stachowiak semiclassical
models, at least at moderate fields and temperatures. In contrast, the 'forbidden orbit' $\beta$ - $\alpha$ is
accounted for by a non-LK equation, involving products of classical amplitudes relevant to $\alpha$ and $\beta$.
Furthermore, strongly non-monotonic field and temperature dependence may be observed for the second harmonics of basic frequencies such as 2$\alpha$ and the MB orbit $\beta$ + $\alpha$, depending
on the value of the spin damping factor $R^s_{\eta}$.

These formulae are in agreement with the dHvA oscillation spectra of the strongly 2D organic metal
$\theta$-(ET)$_4$CoBr$_4$(C$_6$H$_4$Cl$_2$), assuming $g^*$ = 2. However, SdH oscillations spectra exhibit a tremendous number of Fourier components suggesting, besides quantum interference contributions,  more severe deviations from the LK model that remains to be explained within a SdH theory including quantum interference contribution.

\acknowledgments
This work has been supported by EuroMagNET II under the EU contract number 228043, DGES-Spain (Projects FIS2009-12721-C04-03 and CSD2007-00041) and by the CNRS-RFBR cooperation under the PICS contract number 5708. We acknowledge the technical help of Laure Vendier at the X-ray facility of the LCC-Toulouse.

\end{document}